# Ergodic encoding for single-element ultrasound imaging in vivo

Olivier Caron-Grenier[1], Jonathan Porée[1], Vincent Perrot[1], Gerardo Ramos-Palacio[2], Abbas F Sadikot[2] and Jean Provost[1,3]

[1]*Engineering Physics Department, Polytechnique Montréal, Canada, C.P. 6079, Montréal, Québec, Canada, H3C 3A*
[2]*Montreal Neurological Institute, McGill University, Montréal, Canada, H3A 2B4*
[3]*Montreal Heart Institute, Montréal, Canada, H1T 1C8*

Conventional ultrasound imaging relies on the computation of geometric time delay from multiple sensors to detect the position of a scatterer. In this paper, we present Ergodic Relay Ultrasound Imaging (ERUI), a method that utilizes an ergodic cavity down to a single ultrasonic sensor for ultrasound imaging. With the proposed method, the ergodic cavity creates a unique temporal signature that encodes the position of a scatterer. When compared to standard approaches, ERUI enables the generation of images of comparable quality while utilizing fewer detector elements. Our results suggest that ERUI has the potential to achieve image resolution similar to that of traditional imaging techniques, shifting the complexity from hardware to sofware. The demonstrated feasibility offers a promising path towards ultrasound probes with reduced costs and complexity for more portable scanning devices.

Ultrasound imaging is one of the most widely used imaging modalities in clinics due to its capability to image at a large depth, in real-time and without ionization. Typically, ultrasound scanner employs small electrical transducers which emit waves within the 1-40 MHz frequency range, subsequently receiving the echo of the insonified medium. Scatterers and interfaces positions within this medium are retrieved by estimating the delay between the emission and the reception, whereas acoustic impedance of the medium is reflected by the amplitude of the signal. To provide a two-dimensional view of the body's internal structures, ultrasound probes are composed of an array of transducer elements to acquire several delays on a single line. 2D and 3D probes typically require hundred or thousand elements respectively which leads to limitations that have no easily applicable workarounds in both data management and manufacturing in confined spaces. To reduce the element count, compressive sensing and sampling masks have been explored in optics [1–3], photoacoustic [4–6] and ultrasound [7–10] and have even shown the capability to reconstruct a volume with a single ultrasonic element [11]. These techniques require large acquisition times since they involve taking multiple measurements with different configurations of apertures, which slows down the imaging process and post-processing pipeline. To further encode spatiotemporal information, one can turn to chaotic cavities and time reversal acoustic, a topic that has been extensively researched for many years. [12–14] Methods were developed that utilize the spatiotemporal encoding of chaotic cavities to transmit a focused wave to a single location that can be used for high-intensity ultrasound pulses with a limited number of low-power transducer elements [15,16]. By adding a receiving transducer, the backscattered focused wave can also be used to reconstruct an entire image with only one receiving element [17–19]. Although the chaotic cavities developed have important encoding capacities, the pixel-by-pixel focusing in transmit requires long acquisition times. In photoacoustic, a similar approach was developed to encode the received ultrasonic wave by using the encoding of an ergodic relay (i.e. a resonant cavity) [20,21]. This method enables fast acquisition using a small number of detector elements. However, they require recalibration for each new object, which can also be time-consuming. Additionally, the optical diffusion limit [22] inherently bounds photoacoustic imaging at ~1mm depths in biological tissue. Herein we propose Ergodic Relay Ultrasound Imaging (ERUI), which uses an ergodic cavity coupled to a single ultrasonic transducer in receive. Specifically, we show that in combination with plane wave emissions, an ergodic relay positioned to encode only in receive allows for real-time *in vivo* image reconstruction and can achieve high image quality when compared to conventional approaches. With the extension of the calibration methods into the elevation plane, 3D image formation could be readily envisioned.

Based on the ergodic relay described in [20], a right-angle prism made of UV fused silica (PS615, Thorlabs, Inc., 1.5 cm right-angle edge length) was affixed to a 5-MHz linear probe (L7-4, ATL Philips) to be used as a spatiotemporal encoder, as depicted on Fig 1(a). The prism was positioned at one end of the probe and was in contact with 40 (out of 128) transducer elements using polyester resin as a coupling agent, offering the possibility to use any of them as a single-element detector using a programmable ultrafast ultrasound system (Vantage 256, Verasonics). At the opposite side of the probe, 64 elements were used to emit tilted planes waves ($-12:1:12°$) and signals were recorded on all the elements of the probe using a quarter-wavelength, 200% bandwidth sampling scheme, including the ones affixed to the prism (Fig 1.a).

Before imaging, a calibration procedure, consisting in recording the acoustic signature of each pixel of the image to form a dictionary was performed. Briefly, following the theory described in [23], the measurements of piezoelectric elements **y** can be linked to the medium **x** with a linear approximation and can be expressed as :

$$\mathbf{y} = \mathbf{K}\mathbf{x} \qquad (1)$$

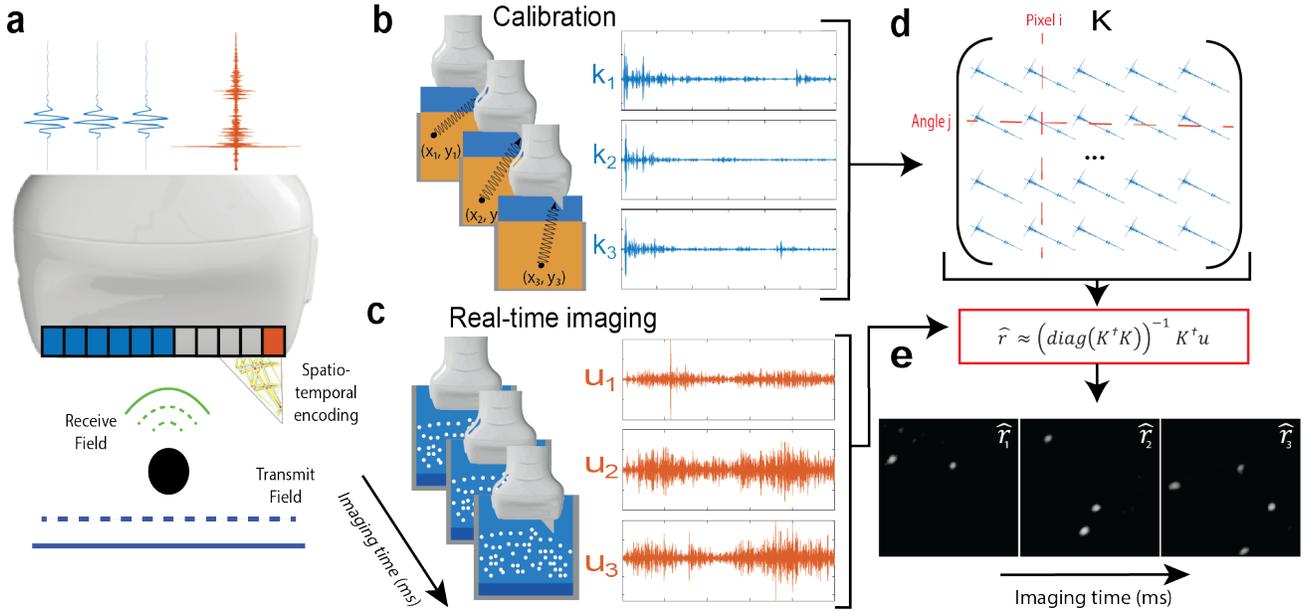

**Figure 1 | Imaging with ERUI.** a) Acquisition of spatiotemporal signals through an ergodic relay. b) During the calibration step, encoded signals from various positions of a unique point source are acquired. This step forms the basis the dictionary K. c) In imaging mode, a plane wave insonifies the entire FOV of the calibration grid at a high frame rate. d) The generated dictionary of calibration 'codas' (**K**), established during the calibration step to reconstruct real-time images. e) Images of free-flowing microbubbles are reconstructed using the dictionary in (d) to decode the signals in real-time.

Where the operator **K** is a collection of projections, i.e.

$$\exists! \, k_m(r), y_m = \langle k_m(r) | x(r) \rangle \quad (2)$$

where $y_m$ is the $m^{th}$ sample of the acquired data of size M and $k_m$ is the associated projection. $k_m(r)$ can be determined, for instance, by setting $x_n = \delta(r - r_n)$, where r and $r_n$ represent the position of a scatterer and the position of a given pixel in the image respectively. This can be performed by discretizing the field of view into *N* single scatterers (typically located at the center of the desired pixels of the reconstructed image) and all the components $k_{m,n}$ of matrix **K** can be determined by repeating the operation for each object containing a single non-zero pixels and denoted by $e_n$. In matrix notation, one would obtain :

$$y = K e_n = k_n \quad (3)$$

Where $k_n$ is the nth line of length M of the matrix K and the entire matrix K can be obtained by repeating the operation for all n. In this case, we have:

$$Y = KI = K \quad (4)$$

Where Y is a matrix containing multiple rows of measurement vectors and I is the identity matrix, i.e., the collection of multiple columns containing canonical vectors $e_i$.

Experimentally, the calibration procedure was performed by recording signals measured by the spatiotemporal encoder from the echo of a 20-um wire in a water tank. We recorded the wave field on a 22 × 22 mm plane located 10 mm in front of the probe using a 3D translation stage (X-LSM200A, Zaber Technologies Inc.) as illustrated in Fig 1(b). The steps size of the calibration grid was 0.15 mm (approximately half the wavelength associated with the emission central frequency) in both the axial and lateral directions. To remove experimental noise from the dictionary, each encoded signal was recorded 30 times and then averaged. Since the received temporal signatures differ for each angled plane wave, a unique signal was obtained for each angle in every pixel, represented as a row in Fig 1(c). Part of the temporal signature was parasitized with signals that were coming directly from the transmitted pulse and reflecting into the prism. To remove those inner reflections, a Singular Value Thresholding was applied to K to eliminate the first 6 values for each dictionary. The calibration procedure required approximately 7 hours, after which the complete spatiotemporal impulse response of the sensor was known. Note that this calibration is in principle specific to the ergodic relay and only needs to be performed once.

Following the forward problem described in (1), the reconstruction problem of finding y from x, can be solved with multiple solvers. Herein, we used a Tikhonov regularization, which consists in limiting the L2 norm of the solution:

$$\hat{x} = \underset{x}{\mathrm{argmin}} \, ||y - Kx||^2 + \alpha ||\Gamma x||^2 \quad (5)$$

Where $||.||$ indicates the L2 norm, $\Gamma$ is a regularization operator and $\alpha$ is a constant.

To enable real-time image reconstruction, and following [23], we applied first iteration the Jacobi algorithm and obtain :

$$\hat{x} = \left(diag(K^\dagger K)\right)^{-1} K^\dagger y \quad (6)$$

Which is approximately equal to the back-projected data divided by the amplitude of the PSF in each pixel.

For real-time imaging procedures, the entire calibration grid is insonified at once and the sum of the backscattered echoes from each scatterer is recorded simultaneously. In our experimental set-up illustrated in Fig. 1(c), the probe was submerged in a 12-liter water tank containing $2.4 \times 10^7$ Definity microbubbles. With 5 tilted plane waves ($-12°:6°:12°$) acquired at a frame rate of 250-Hz and a dictionary **K** of 125×125 pixels, the reconstruction of the moving microbubbles in Fig. 1(e) and Supplementary Video 1, was performed in 2.3 ms per frame.

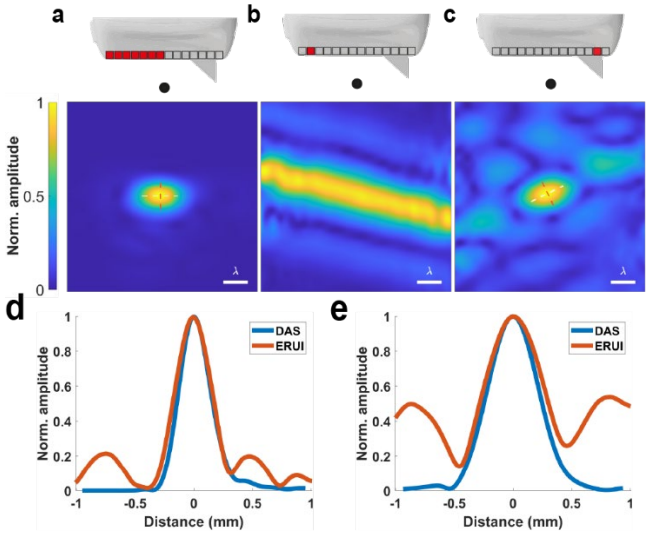

**Figure 2 | Reconstruction of a thin wire using various reconstruction methods.** a) Delay-and-Sum (DAS) algorithm using the same elements used for transmitting the emission pulse. b) Spatiotemporal matrix image formation (SMIF) where the matrix K was defined by the measurement of a single element not covered by the ergodic relay. c) Ergodic relay ultrasound imaging (ERUI) reconstruction using a single element covered by the ergodic relay. d) Axial profile of DAS and ERUI taken along the red dotted line. e) Lateral profile of DAS and ERUI taken along the white dotted line.

To compare ERUI reconstruction to typical imaging methods, a single wire was imaged (Fig. 2) with delay-and-sum (DAS), spatiotemporal matrix image formation or SMIF (i.e., the reconstruction described in (6) with a single element not covered by the ergodic relay) and ERUI. First, DAS beamforming was performed on demodulated channel data, on a 2.4×2.4 mm cartesian grid using an in-house GPU implementation of the delay and sum beamformer [24]. Using the DAS image as a benchmark, we observed in Fig. 2(b)-(c) that ERUI can reconstruct the scatterer, while in Fig. 2(b) SMIF method with only one element fails, as expected, to accurately localize the scatterer. Fig. 2(d)-(e) provide axial and lateral profiles respectively, comparing DAS and ERUI. Resolution measured at -10dB is comparable between ERUI and DAS with a lateral resolution of $2.4\lambda$ and $2.1\lambda$ and an axial resolution of $1.5\lambda$ and $1.4\lambda$ respectively.

To assess the method in a more realistic setup a 1450 m/s ATS 539 Multi-purpose phantom (ATS Laboratories, BT, USA) was used. The 0.12-mm diameter wires used to measure resolution were located between 25- and 30 mm depths from the probe. 25 plane waves with a fixed sampling of 1° were emitted. A comparison of DAS and ERUI with all 25 angles in Fig. 3(a) shows that multiple wires can be effectively imaged using either method. The resolution of the reconstructed wire with a lower number of transmission angles was quantified in Fig. 3(b)-(c) by measuring the peak width at $-10$ dB of one of the wires. While a wire in the presence of speckle can be observed with a single plane wave, the compounding of multiple angles quickly improves the lateral resolution of the reconstruction.

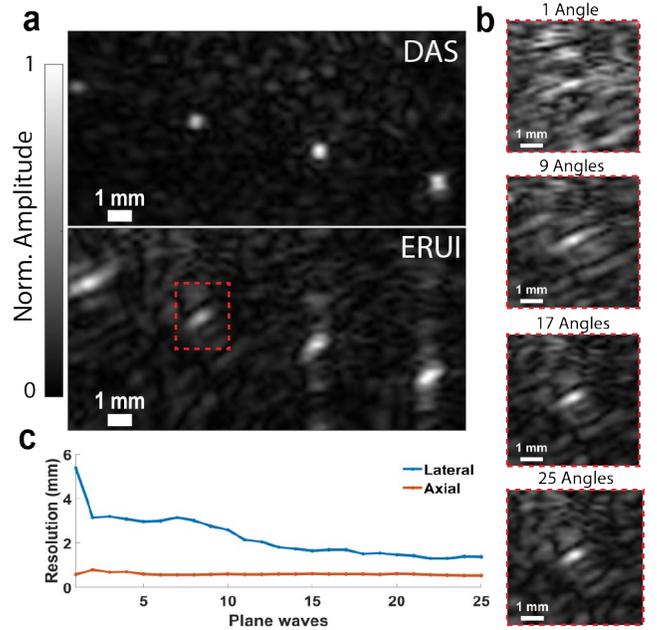

**Figure 3 | Quantification of resolution in phantom** a) Comparison of the reconstruction of multiples targets from a commercial imaging phantom with DAS and ERUI. The upper image shows a DAS reconstruction using the 64-transmission element while the lower image uses ERUI, with both methods using 25 compounded plane waves spaced by 1°. b) Close-up view of the ERUI reconstruction of the red box in (a) with an increased number of compounding angles. c) Variation of the measured -10 dB resolution of the target illustrated in (b) with multiples transmitted angles.

Contrast is typically measured with an anechoic inclusion, as depicted with typical image formation in Fig. 4 (a). Here contrast was measured by evaluating the overlap of pixels in the anechoic region ($p_{in}(x)$) and in the background ($p_{out}(x)$) using the generalized contrast-to-noise ratio as described in [25]:

$$\text{gCNR} = 1 - \int_{-\infty}^{\infty} \min_x p_{out}(x), p_{in}(x) dx \qquad (7)$$

Using a method based on the acquisition of long reverberant signals, the detection of anechoic inclusion (i.e. a region devoid of signals) to measure contrast poses a certain challenge. As shown in Fig. 4(c), with a single element and single or multiple emission angles, the inclusion is not easily detectable. To achieve a better signal-to-noise ratio, an approach with multiple dictionaries, each associated with a specific element, was used. In this approach, the multiples reconstructions obtained with different dictionaries were combined to form a final image. For the imaging process shown in Fig. 4(c), a single element and a total of 10 elements covered by the ergodic relay were used. The same process was carried out with the elements located at the other end of the probe and thus not covered by the ergodic element in Fig. 4(b). For both methods, as the number of elements and angles employed increased, a noticeable enhancement in the distinction of the target region became evident, both visually and via gCNR measurements. In the case of anechoic inclusion, the addition of an ergodic relay appears to degrade contrast when compared to a set-up without one.

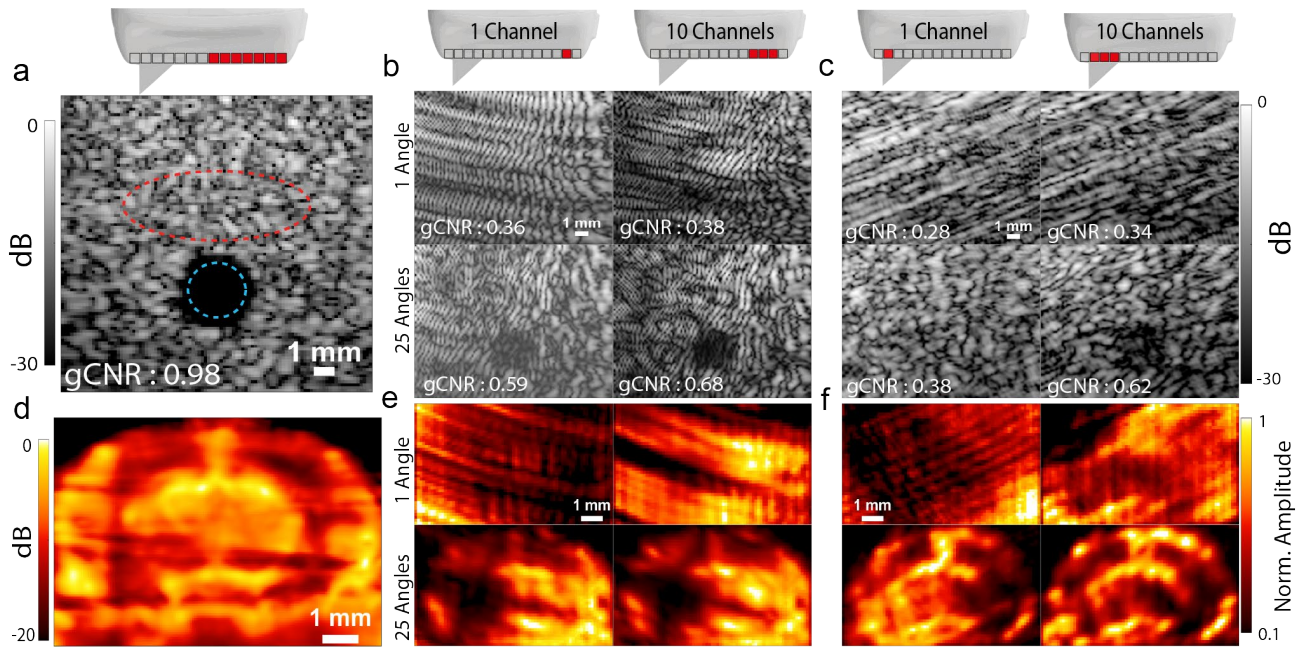

**Figure 4 | Reconstruction of anechoic cyst and a mouse brain.** a) Reference image of an anechoic region formed with delay and sum. The anechoic region used for all gCNR measurements is outlined in blue and the background region in red. b) and c) Anechoic inclusion imaged with SMIF and ERUI respectively. For each method, the region is reconstructed with a single receive element (left) and ten compounded receive element (right) while using a single transmission angle (top) and 25 transmission angles (bottom). d) Reference power Doppler of a mouse brain with delay and sum used as the imaging method. e) and f) Power Doppler of the mouse brain with images formed with SMIF and ERUI respectively and expressed in normalized amplitude.

In vivo acquisitions were performed transcranially on the brain of an eight-week-old wild-type female mouse anesthetized with Isofluorane (2%) and placed on a stereotaxic fixation system, per McGill University Animal Care Committee regulation under protocol #2001-4532. As better results were obtained in a medium with a reduced number of strong emitters and with a high SNR, a microbubble solution (Definity, Lantheus Medical Imaging) was injected containing 4μL/g microbubbles diluted in a 1:10 ratio with saline. To obtain a Doppler image, 40 ensembles of 250 frames each were acquired with a frame rate of 250 Hz. Reconstruction of the mouse brain was obtained with DAS, SMIF and ERUI in Fig. 4(d), (e) and (f) respectively. Tissue signal suppression was achieved by applying a SVD thresholding [26] and setting the first 15 singular values to zero.

For the 5-MHz emission frequency of the probe, a mouse brain through the skull is not an ideal model to illustrate Doppler imaging capability, as only the main vessels are detected. However, with the detection of sparser scatterers in the form of microbubbles, our results tend to show that ERUI can reconstruct anatomical structure with a single element and with better resolution than SMIF.

In this Letter, we described a novel approach of ultrasound images reconstruction, characterized by the reduced number of elements required by encoding spatiotemporal information via an ergodic relay. The findings validate the potential of an ergodic cavity to encode ultrasound codas and perform ultrasound imaging with a single or minimal set of sensors which deviates from the conventional requirement. ERUI draws heavily on the foundational ideas proposed by various researchers within the field of time-reversal acoustics and imaging via compressed sensing. The novelty of our approach is in the encoding of information in receive only. This shift optimizes insonification of scatterers while preserving the complexity of the backscattered signal, thereby enabling real-time imaging with a simple yet effective algorithm for reconstruction.

The efficacy of single-element imaging apparatus often hinges on their capability to form a complete orthogonal basis utilizing constrained information resources. This capability is intrinsically linked to the successful reconstruction of a region with multiple scatterers. In this context, our findings suggest a promising, though not yet unequivocal evidence supporting this goal, with discernible improvement in imaging resolution with a certain cost in contrast. However, given impedance mismatch present in our experiments between the medium and the ergodic relay stemming from the lack of direct contact between the two, the increase of image quality with a better SNR gives us confidence in the potential of the proposed system under enhanced operational conditions. Of particular interest is the current calibration process, which employs water as the propagation medium. Given that this typically does not correspond with the speed of sound employed in conventional ultrasound imaging, the capability to alter the calibration medium in future apparatus iterations could offer significant room for improvement. Additionally, the calibration method restricts the reconstruction to a two-dimensional plane, but we foresee the potential for implementing a three-dimensional calibration method in future iterations of the apparatus.

In summary, with the ergodic encoding of ultrasound signals we believe that ERUI will pave the way for an entirely new means of imaging in which the complexity is further shifted away from hardware and to software.

We acknowledge the support of FRQNT, TransMedTech, IVADO, CIHR, NSERC (DGECR-2020-00229) and of the CFI (38095 and 246916). This research was enabled in part by support provided by Calcul Quebec (calculquebec.ca) and the Digital Research Alliance of Canada (alliancecan.ca). (Corresponding author: Jean Provost).